\documentclass[aps,preprint,superscriptaddress,amsmath,amssymb]{revtex4}

\usepackage[dvips]{graphicx}
\usepackage{dcolumn}
\usepackage{amsmath}
\usepackage{threeparttable}

\usepackage{bm}
\usepackage{bbm}

\usepackage[version=3]{mhchem} 

\newcommand{\be}{\begin{equation}}
\newcommand{\ee}{\end{equation}}
\newcommand{\bea}{\begin{eqnarray}}
\newcommand{\eea}{\end{eqnarray}}
\newcommand{\bean}{\begin{eqnarray*}}
\newcommand{\eean}{\end{eqnarray*}}

\begin{document}


\title{Algorithmic differentiation and the calculation of forces by quantum Monte Carlo}

\author{Sandro Sorella}
\email{sorella@sissa.it}
\affiliation{ SISSA, International School for Advanced Studies, 34151, Trieste, Italy}
\affiliation{DEMOCRITOS National Simulation Center, 34151, Trieste, Italy}

\author{Luca Capriotti}
\email{luca.capriotti@credit-suisse.com}
\affiliation{Quantitative Strategies, Investment Banking Division, Credit Suisse Group, 
Eleven Madison Avenue, New York City, NY 10010-3086, United States of America} 

\date{\today}

\begin{abstract}
We describe an efficient algorithm to compute forces in quantum Monte 
Carlo using adjoint algorithmic differentiation.
This allows us to apply the space warp coordinate transformation in differential form,
and compute all the $3M$ force components of a system with $M$ atoms with a 
computational effort comparable with the one to obtain the total energy.
Few examples illustrating the method for an electronic system containing  
several water molecules are presented. With the present technique, 
the calculation of finite-temperature thermodynamic properties 
of materials with quantum Monte Carlo will be feasible in the near future. 
\end{abstract}

\maketitle

\section{Introduction}
\label{introduction}
In the last few years, we have seen 
 a remarkable progress in the ab-initio simulation 
of realistic electronic systems based on first principles quantum 
mechanics.
Despite the power of density functional theory (DFT), with standard 
local density approximation (LDA) and generalized gradient approximation (GGA) functionals, 
much effort has been devoted to schemes 
that are able to describe more accurately the electronic correlations.
This is because several materials -- such as high temperature superconductors -- 
are indeed strongly correlated. Furthermore, long-range 
dispersive forces 
may be extremely important even in simple and fundamental materials, like water, 
and are notoriously difficult to describe with standard DFT.
A promising many-body approach, alternative to DFT, 
is the so called quantum Monte Carlo (QMC) method,
allowing one to include the electronic correlations by means of a highly-accurate 
many-body wave function (WF), sampled by a statistical method.
All the basic ingredients of the electronic 
correlation are described explicitly  within  this framework.  This is 
particularly appealing because its computational cost scales 
rather well with the number of electrons $N$, with a modest power, e.g., $N^3$ or $N^4$. 
Due to this important property, QMC is very promising for large scale calculations especially 
when compared with standard post-Hartree-Fock methods. In fact, these methods are 
also capable to describe rather well the electronic correlations. However, they typically require a larger 
computational cost: from polynomial in the range between $N^4$ and 
$N^7$, to exponential complexity in full configuration interaction (FCI) schemes.
     
Despite the clear advantage of QMC for the accurate electronic 
simulation of materials containing a large number of atoms, its application 
has been mainly limited to total energy calculations.
In particular, no general method to calculate ionic forces, that remains efficient even  for a large number of atoms $M$, is available so far.
Indeed, many-body forces usually lead to cumbersome expressions,
whose evaluation can be done at present only by means of complicated and computationally expensive  algorithms.  For this reason,  such calculations have been implemented so far only for particularly simple cases.
For instance, it was recently possible \cite{attacc} to simulate $ \sim 100$ 
hydrogen atoms in the low temperature, high pressure phase, with the 
calculation of the ionic forces based on 
efficient  strategies to work with finite variance expressions \cite{caffarell,caffforce}. 
Unfortunately, this route cannot be followed in general because it works very efficiently 
only for light atoms.
 
On the other hand, an  accurate algorithm for the calculation 
of forces 
 that is in principle 
efficient also with heavy atoms, and with the use of pseudopotentials,  
was introduced a long time ago \cite{warp}.
This method is based on the so called
the {\em space warp coordinate transformation} (SWCT), allowing a zero-variance expression -- i.e., maximum efficiency in QMC --  
even for isolated atoms.
We believe that, without a highly effective variance reduction scheme in the calculation 
of forces,  such as SWCT, there is no hope  to extend the applicability of QMC to structural optimization of complex and correlated 
materials, or to perform  ab-initio molecular dynamics simulation at finite 
temperature. However, even when using this promising approach, or others  
based on the zero-variance principle \cite{caffforce}, 
standard implementations, e.g.,
based on finite-difference approximations of the derivatives appearing in the
expressions for the ionic forces,  are still computationally very expensive -- to the
point of being unfeasible for a large number of atoms.

In this work, we propose a simple strategy for the 
efficient calculation of the ionic forces
 -- and generally of any  arbitrarily complicated derivative of the 
 QMC total energy --
by using {\em adjoint algorithmic differentiation}. 
We will show that this method will allow us to achieve two very 
important targets:
\begin{enumerate}
\item  the numerical implementation of  all the energy derivatives will be possible in a straightforward  way without any reference to complicated expressions, as for instance the terms 
shown in Ref.~\onlinecite{needs}, when pseudopotential are used to 
remove the effect of  core electrons;

\item more importantly, the calculation of {\em an arbitrarily large} 
number of energy derivatives will be possible with a computational effort 
comparable with the one to compute the energy alone. In other words, once 
the calculation of the energy is well optimized, by  means of AAD, also the
 one for the energy derivatives will be almost optimal. 
\end{enumerate}
The latter  property is rather remarkable, and suggests that most of 
the  advanced ab-initio tools belonging to a rather restrictive approximation 
of the DFT functional, such as structural optimization and molecular 
dynamics at finite temperature, can be extended to highly accurate many-body  
approaches based on QMC, with a computational effort that remains 
 affordable even for a large number of atoms.

Though in most of the examples presented here we have used the standard variational QMC 
method, the technique we propose in this paper directly
applies also to more accurate QMC projection 
schemes, such as (lattice regularized) diffusion Monte Carlo. 
However one has to take into account that in this case there are unsolved  
technical  problems -- e.g., infinite variance in the most 
accurate  estimators -- that 
do not depend on the technique we propose, and are  
outside the main  scope of this paper. 
In order to avoid confusion, we anticipate that we apply the AAD 
technique only to the  specific calculation of the wave function and 
the local energy (see later for their definitions)  
required for the VMC or DMC (LRDMC)  evaluation of the average energy. In 
principle AAD can be applied to the whole algorithm, but we have 
not exploited this possibility, that may be interesting for future applications.  
\section{QMC wave function, VMC and LRDMC }

In this Section we begin by describing the WF that we have used in our QMC calculations.
In the following we will denote with $\mathbf{ r} $ a generic three dimensional
electronic position, whereas $x=\{ \mathbf{r}_1, \mathbf{r}_2, \cdots ,
 \mathbf{r}_N \} $ stands for a configuration of
all electron positions and spins;  the $N_\uparrow$ spin-up
electrons are at positions $\mathbf{r}_i$ with $1\le i \le N_\uparrow$ , and the
$N_\downarrow$ spin-down electrons are at positions $N_\uparrow +1 \le i \le N $.

The usual trial WF used in the QMC calculation is the product of an
antisymmetric part, and a Jastrow factor. The antisymmetric part is a single Slater determinant, while the Jastrow factor is a bosonic 
many-body function which accounts for the dynamic correlations 
in the system. Our Slater determinant is obtained with
$N/2$  doubly occupied molecular orbitals $\psi_j(\mathbf {r} )$, expanded over  $L$ atomic Gaussian 
orbitals $\phi_j( \mathbf{ r})$, centered at atomic positions $\mathbf{R}_j$, 
 as \cite{marchi}:
\begin{equation} \label{defpsi}
\psi_i( \mathbf{ r} ) = \sum \limits_{j=1}^L \chi_{ij} \phi_j( \mathbf{ r}) 
\end{equation}
where the coefficients $\chi_{i,j}$, as well as the non-linear coefficients, 
appearing in the exponents of the Gaussians, 
can be fully optimized by energy minimization as described later.
 
The molecular orbitals are initialized  from a  
 self consistent DFT-LDA calculation,  in the same atomic basis. 
 The Jastrow factor takes into account the electronic correlation between two 
electrons, and  is conventionally split into an homogeneous interaction $J_2$, depending on the relative 
distance between two electrons,  and two non-homogeneous contributions $J_3$ and $J_4$, 
depending on the positions of two electrons and one atom, and two electrons and two atoms, respectively. 
It also contains an inhomogeneous term $\tilde J_2$, describing the electron-ion interaction. This 
is important to compensate for the 
change in the one particle density induced by $J_2$, $J_3$ and $J_4$, as well as to satisfy the electron-ion cusp conditions.
 The homogeneous and inhomogeneous  two-body terms $J_2$ and $ \tilde J_2$ are defined by the following equations: 

\begin{eqnarray}
\label{j1}
\tilde J_2=\exp{\big[\sum_{ia}-(2Z_a)^{3/4}u(Z_a^{1/4}r_{ia})+\sum_{ial}
g_l^a \chi_{al}^J(\mathbf{r}_i)\big]},
\end{eqnarray}
and 
\begin{eqnarray}
\label{j2}
J_2=\exp{[\sum_{i<j}^{}u(r_{ij})]},
\end{eqnarray}
where $i,j$ are indices running over the electrons, and $l$ runs over different single particle orbitals $\chi_{al}^J$ 
centered on the atomic center $a$; $r_{ia}$ and $r_{ij}$ denote 
 the electron-ion and the 
electron-electron distances, respectively. The corresponding cusp conditions are fixed by 
the function $u(r)=F[1-\exp(-r/F)]/2$ (see e.g., Ref.~\cite{rocca}), whereas 
$g_l^a$ and $F$ are optimizable variational parameters. 

The three and four-body Jastrow terms $J_{3} J_{4}$ are  given by:
\begin{equation}
\label{jastrow}
J_{3}(x)  J_{4}( x)=
\exp\left(\sum \limits_{i<j} f(\mathbf{ r}_i ,\mathbf{ r}_j)\right), 
\end{equation}
with $f (\mathbf{ r}_i ,\mathbf{ r}_j)$, being  a two-electron coordinate 
function that  can be expanded into the same single-particle basis 
used for $\tilde J_{2}$: 
\begin{eqnarray}\label{3bjsp}
f(\mathbf{r}_i,\mathbf{r}_j)=
\sum_{ablm}^{} g_{lm}^{ab}\,\chi_{al}^{J}(\mathbf{r}_i)
\chi_{bm}^{J}(\mathbf{r}_j),
\end{eqnarray}
with $g_{lm}^{ab}$ optimizable parameters. Three-body (electron-ion-electron) correlations are described by the diagonal matrix elements 
$g^{aa}$, whereas four-body correlations (electron-ion-electron-ion) are described by the 
matrix elements with $a\ne b$.

The  complete expression of the  Jastrow factor  $J( x) = J_2(x) \tilde J_2 (x) J_3(x) J_4(x) $ 
that we adopt in this work allows us to take into account weak and long-range electron-electron interactions, 
and it is also extremely effective for suppressing  higher energy configurations occurring when electrons are too close.  
  
In order to minimize the energy expectation value corresponding to this WF 
we have applied the well-established  energy minimization 
 schemes \cite{casulamol,rocca,umrigar}, that 
we have recently adapted for an efficient optimization of the molecular
 orbitals of the Slater determinant in presence of the Jastrow factor described above \cite{marchi}.

In variational Monte Carlo (VMC), the energy expectation value of a given 
correlated WF, depending on a set of variational parameters 
$c_i, i=1,\cdots p$, can be computed with a standard statistical method.
The energy depends in turn on the atomic positions ${\mathbf{R}_a}$, 
$a=1, \cdots M$, so that we indicate formally:
\begin{equation}
E(\{c_i\},\{\mathbf{R}_a\})= { \langle \Psi_{\{c_i\},\{\mathbf{R}_a  \} }
 |  \hat H_{\{\mathbf{R}_a  \}} | \Psi_{\{c_i\},\{\mathbf{R}_a  \}} \rangle
 \over 
 \langle \Psi_{\{c_i\},\{\mathbf{R}_a  \}}
 | \Psi_{\{c_i\},\{\mathbf{R}_a  \} } \rangle }~.
\end{equation}
In VMC the above energy expectation value is computed statistically by sampling the probability $ \Pi (x) \propto  \langle x | \Psi \rangle^2 $.
Analogous and more accurate techniques are also possible within QMC.
In this work the lattice regularized diffusion Monte Carlo (LRDMC) will be also used. 
The latter method is  a projection 
technique, filtering out the ground state component of a given 
variational WF by applying the propagator $ e^{ - H \tau } $ 
to $ \Psi$ for large imaginary time  $\tau $. This propagation is 
subject to the 
restriction to modify only the amplitudes of the WF without 
 affecting  its phases. 
In this way one can avoid the so called ``fermion sign problem'' 
instability in QMC, with an highly  accurate technique,  
providing a rigorous upper bound of the total energy even in presence of 
pseudopotentials \cite{lrdmc}.

\section{ Space warp coordinate transformation and its differential form}
 
The  main purpose  of this Section is to describe an efficient  method
  to compute 
the  forces $\mathbf{F}$ acting on  each  of the  $M$ nuclear positions
$\{\mathbf{R}_1, \ldots , \mathbf{R}_M\}$, namely,
\be
\label{force}
\mathbf{F}(\mathbf{R}_a)=-\mathbf{\nabla}_{\mathbf{R}_a}
 E(\{c_i\}, \{\mathbf{R}_a\}),
\ee
with a reasonable statistical accuracy. 
Following Ref.~\onlinecite{casulamol}, we introduce
a finite-difference operator $ { \Delta/\Delta\mathbf{  R}_a } $ for the  evaluation of the 
force acting on a given  nuclear position $\mathbf{R}_a$, 
\begin{equation} \label{forcefinite}
{ \Delta \over {\Delta \mathbf{R}_a} }E =
 { E(\mathbf{R}_a + \Delta\mathbf{ R}_a )  -  E(\mathbf{R}_a )    
  \over  {\Delta \mathbf{ R}_a}   } 
\end{equation}
so that
\begin{equation} 
\mathbf{F} (\mathbf{R}_a)=- { \Delta \over \Delta \mathbf{ R}_a }E + O({\Delta R}) 
\end{equation}
where $\Delta \mathbf{R}_a$ is a three dimensional vector.
Its length $\Delta R $ can be  chosen as small as  $10^{-6}$ atomic 
units, yielding negligible finite-difference errors for the evaluation of the exact energy derivative.

In order to evaluate the energy differences in Eq.~(\ref{forcefinite}) 
it is very convenient to apply 
 the space warp coordinate transformation (SWCT).
This  transformation 
 was introduced a long time ago in Ref.~\onlinecite{warp}, 
for an efficient 
calculation of the ionic forces within VMC.
According to this transformation, as soon as the ions are displaced, 
 also the electronic coordinates 
$\bf r$ will be translated in order to mimic
the displacement of the charge around the nucleus  
$\mathbf{R}_a$, namely $x \to \bar x$, with
\be \label{spacewarp}
\overline{ \mathbf{r}}_i=\mathbf{r}_i+
\Delta\mathbf{ R}_a ~ \omega_a(\mathbf{r}_i),
\ee
\be
\omega_a(\mathbf{r})=\frac{F(| \mathbf{r}-\mathbf{R}_a|)}
{\sum_{b=1}^{M} F(| \mathbf{r}-\mathbf{R}_b|)}~,
\ee
and $F(r)$ is a function which must decay rapidly; here we used 
$F(r)={1}/{r^4}$ as suggested in Ref.~\onlinecite{filippi}.

The expectation value of the energy depends on
$\Delta\mathbf{ R}_a$, because both the Hamiltonian and the WF depend
on the nuclear positions. Applying the SWCT to
the integral involved in the calculation, the expectation value reads
\be \label{forcewarp}
E( \mathbf{R_a}  +\Delta\mathbf{R}_a)=
\frac{\int d\mathbf{r}^{3N}  \tilde J_{\Delta\mathbf{R}_a}(x)
\Psi_{\Delta\mathbf{R}_a}^2 (\bar x (x))
E^{\Delta\mathbf{R}_a}_L(\bar x (x))}
{\int d\mathbf{r}^{3N}  \tilde J_{\Delta\mathbf{ R}_a}(x) 
\Psi^2_{\Delta\mathbf{ R}_a}(\bar x (x))},  
\ee
where $\tilde J$ is the Jacobian of the transformation and,  
\begin{equation} \label{elocal}
E_L^{\Delta\mathbf{ R}_a}= 
{ \langle \Psi_{\Delta\mathbf{R}_a} | \hat H |x \rangle \over \langle \Psi_{\Delta\mathbf{ R}_a} |x \rangle } 
\end{equation}
 is 
 the so called {\em local energy} defined 
 by the wave function 
 $\Psi_{\Delta\mathbf{ R}_a}$
on a real space  electronic configuration $x$. 
In the following, we define 
$E_L=E_L^{\Delta\mathbf{ R}}$ for ${\Delta\mathbf{ R}_a}=0$.

The importance of the SWCT in reducing the 
statistical error in the evaluation of the force is easily understood 
for the case of an isolated 
atom $a$. In this case, the force acting on the atom is obviously 
 zero, but only after the SWCT 
with $\omega_a=1$ the integrand  
in Eq.~ (\ref{forcewarp}) is independent of $\Delta\mathbf{ R}_a$, 
providing an estimator of the force with zero variance. 

Starting from Eq.~(\ref{forcewarp}), 
it is straightforward to  derive explicitly a  
differential expression for the force estimator, 
which is related to the gradient of the previous quantity with respect
to $\Delta\mathbf{ R}_a$ in the limit of vanishing displacement, 
\begin{eqnarray}
\label{vmcforce}
\mathbf{F}_a 
 & = & - \big \langle  
 \frac{d }{d \mathbf{R}_a} E_L \big \rangle  
\\ \nonumber
& + &  2 \Big ( 
\big \langle  E_L \big \rangle \big \langle
\frac{d}{d \mathbf{ R}_a}
\log (J^{1/2}  \Psi  ) \big \rangle -
\big \langle E_L   
\frac{d }{ d \mathbf{ R}_a}
\log (J^{1/2}  \Psi  ) \big \rangle 
\Big ),
\end{eqnarray}
where the brackets indicate a Monte Carlo like average over the 
square modulus of the trial WF, namely over 
the probability $\Pi (x)$ 
introduced in the previous Section.
In  the calculation of the total derivatives $ { d  / d\mathbf{R}_a} $, we have to take into account 
that the electron coordinates are also implicitly differentiated, according to the SWCT. Then, all the  terms above 
can be written in a closed expression once the partial derivatives of the local energy and of the 
WF logarithm are known, namely,
\begin{eqnarray}
{ d \over d  \mathbf{ R}_a } E_L &=& {\partial \over \partial  \mathbf{ R}_a } E_L + \sum \limits_{i=1}^N 
 \omega_a (\mathbf{ r}_i)  { \partial \over \partial \mathbf{ r}_i } E_L~,  \label{locdiff} \\
\frac{d}{d \mathbf{ R}_a} \log (J^{1/2}  \Psi  ) &=& {\partial \over \partial  \mathbf{ R}_a }   \log ( \Psi) + \sum \limits_{i=1}^N \left[  
 \omega_a ( \mathbf{  r}_i)  { \partial \over \partial \mathbf{ r}_i }  \log \Psi + 
{ 1 \over 2 } { \partial \over \partial \mathbf{ r}_i }  \omega_a (  \mathbf{ r}_i) \right]~, \label{puldiff} 
\end{eqnarray}
where the  term  ${ 1 \over 2 } { \partial \over \partial \mathbf{ r}_i }  \omega_a (  \mathbf{ r}_i)$ in the square brackets gives the contribution of the 
Jacobian.

Based on the expressions above we can evaluate the forces in Eq.~(\ref{vmcforce}) 
in three different ways, listed below in increasing order of efficiency:
\begin{enumerate}
\item Helmann-Feynmann (HFM). By neglecting all the dependence of $\Psi$ on the 
atomic position, and without the SWCT: 
 $$\mathbf{F}_a  =- \langle  \partial_{\mathbf{R}_a} \hat H  \rangle~. $$
This is the least accurate expression, because without the so called Pulay
terms derived in Eq.~(\ref{puldiff}) the force is consistent with the energy derivative
only when the WF is the exact ground state. 
This is also the least efficient way to compute the forces as indicated in 
Tab.~\ref{tabeff}.

\item No-SWCT. This is obtained by using the terms (\ref{locdiff},\ref{puldiff}) in 
Eq.~(\ref{vmcforce})  with $\omega_a ( \vec r ) =0$.
This expression is much more accurate and efficient than the previous one,
as it fulfills the so called ``zero-variance principle'':
if the WF $\Psi$ coincides with the  exact ground state for arbitrary 
atomic position $\mathbf{R}_a$, it is easy
to realize that the estimator of the forces 
does not have statistical fluctuations, as the local energy and its derivative 
are just constant, and independent of the real space configuration $x$.

\item Differential SWCT. The properties above are clearly fulfilled 
also in this case.
Moreover, 
the SWCT does not change the mean value of the forces, but 
affects only their statistical fluctuations.
In fact,  as mentioned before, the SWCT allows us to obtain  
a zero-variance property 
for the forces acting on isolated atoms.  As a result, this transformation is
 extremely important for computing forces between atoms at large distance, 
as in this limit they can be considered isolated. 

\end{enumerate}

The advantage to use the SWCT for a water dimer molecule 
is illustrated in Tab.~\ref{tabeff}. It is clear that without the SWCT ,
or even without the differentiation of the local 
energy with respect to the atomic position (no-SWCT case), the evaluation of forces with a reasonable statistical error is simply not possible.  

As discussed in the following Section, all partial derivatives involved 
in the expressions above, namely the $6N +6M$ components,  
$  { \partial \over \partial \mathbf{ r}_i } E_L,
 { \partial \over \partial \mathbf{ r}_i } \log \Psi,  {\partial \over \partial  \mathbf{ R}_a } E_L,  {\partial \over \partial  \mathbf{ R}_a } \log \Psi $,  
 can be evaluated very efficiently with algorithmic 
differentiation. This is true also when the WF and the expression for the local 
energy are extremely cumbersome, e.g., when using pseudopotentials.
As a result,  the quantities in  Eqs.~(\ref{locdiff},\ref{puldiff}) can be evaluated 
using  just a minor computational effort with roughly  $ \propto  N M $ operations. 
In particular, one of the most involved contribution to the local energy is the one 
corresponding to the bare 
kinetic energy $ \hat K= -{ 1\over 2} \sum \limits_{i=1}^N 
\Delta_i $. The Hamiltonian $\hat H$ in Eq.~(\ref{elocal}) always contains this 
term, even in presence of pseudopotentials.
 In the following, we will discuss how to differentiate the  contribution 
 $K = { \langle \Psi | \hat K | x \rangle / \langle \Psi | x \rangle } $ 
to the local energy for the particularly simple but instructive case of 
a Slater determinant WF with no Jastrow factor.  

\section{Adjoint Algorithmic Differentiation}
\label{adsec}

Algorithmic differentiation (AD) \cite{griewank} is a set of programming
techniques for the efficient calculation of the derivatives
of functions implemented as computer programs.
The main idea underlying these techniques is the fact that any such function
 -- no matter how complicated -- can be interpreted as the composition of
more elementary functions each of which is in turn a composition of
basic arithmetic and intrinsic operations that are easy to differentiate.
As a result, it is possible to calculate the derivatives of the
outputs of a program with respect to its inputs by
applying mechanically the rules of differentiation --
and in particular the {\em chain rule} -- to the composition of
its constituent functions. 

What makes AD particularly attractive, when compared to standard
(finite-difference) methods for the calculation of derivatives is
its computational efficiency. In fact, AD exploits the information
on the calculations performed by the computer code, and the
dependencies between its various parts, in order to optimize the
calculation. In particular, when one requires the derivatives
of a small number of outputs with respect to a large number of inputs,
the calculation can be highly optimized
by applying the chain rule through the instructions of the program
in opposite order with respect to the one of evaluation of the
original instructions. This gives rise to the so called adjoint
(mode of) algorithmic differentiation (AAD).

Even if AD has been an active branch of computer science for
several decades, its impact in other research fields has
been surprisingly limited until very recently \cite{autodiff}.
Only over the past two years its tremendous 
effectiveness in speeding up the calculation of sensitivities 
e.g., in Monte Carlo simulations, has been first exploited
in computational finance applications \cite{caprio1}. In particular, the potential
of AD has been largely left untapped
in the field of computational physics where, 
as we demonstrate in the following, 
it could move significantly the boundary of what can be studied 
numerically with the computer power presently available. 

Griewank \cite{griewank} contains a detailed discussion
of the computational cost of AAD \cite{caprio2}.  Here, we will only recall the main
ideas underlying this technique to clarify how it can be beneficial in 
the implementation of the calculation of the forces in QMC. To this end, 
we consider a particular computer implemented function $ X \rightarrow Y $ 
\begin{equation}\label{function}
Y = \texttt{FUNCTION}(X)
\end{equation}
mapping a vector $X$ in $\mathbb{R}^n$ in a vector $Y$ in ${\mathbb R}^m$ 
through a sequence of two sequential steps
\[
X\ \rightarrow\  U\ \rightarrow\ V\ \rightarrow\  Y.
\]
Here, each step can be a distinct high-level function, or even an individual instruction in a computer code. 
A general code is usually implemented by several steps of this type, and, 
more importantly,  the output of a particular instance can be used
as an input  not only for  the next one   
but generally for all instances  occurring later in the algorithm.
Generally speaking an algorithm can be viewed as a sequential 
tree or graph with connectivity larger than one, where each node has more than one children. However,
to keep things as simple as possible, in this ``warm up'' example we do not consider 
these more complex cases. The generalization to a realistic computational graph
is however straightforward \cite{caprio2}.

The adjoint mode of AD results from propagating the derivatives of the final result
with respect to all the intermediate variables -- the so called {\em adjoints} --
until the derivatives with respect to the independent variables are formed.
Using the standard AD notation, the adjoint $ \bar V $ 
of any input  variable $V$  
of an instance $ V \rightarrow Y$ is defined as the derivative 
of a given  
 linear combination of the output $ \sum \limits_j \bar Y_j Y_j$
with respect to the input $V$, namely:
\begin{equation} \label{adjointdef} 
\bar V_k = \sum_{j=1}^m \bar Y_j \frac{\partial Y_j}{\partial V_k} ~,
\end{equation}
where $\bar Y$ is a given input vector in ${\mathbb R}^m$.
In particular, for each of the intermediate variables, using the chain rule, we get,
\[
 \bar Y_j \frac{\partial Y_j}{\partial X_i} =    \bar Y_j  
  \frac{\partial Y_j}{\partial V_k}   \frac{\partial V_k}{\partial U_l}
  \frac{\partial U_l}{\partial X_i}
\]
where repeated indices indicate implicit summations.
It is simple to realize that in such a simple case we can use the definition 
in Eq.~(\ref{adjointdef}) to evaluate $ \bar X$, namely:
\[
 \bar Y_j \frac{\partial Y_j}{\partial X_i} =    \bar V_k   
   \frac{\partial V_k}{\partial U_l}
  \frac{\partial U_l}{\partial X_i} = \bar U_l \frac{\partial U_l}{\partial X_i}  = \bar X 
\]
In other words, once all adjoint instances have been defined, the bar input 
of each adjoint instance can be obtained from the output of the previous 
adjoint instance according to a diagram that follows very straightforwardly 
the original algorithm in reversed sequential order:
\begin{equation}
\bar Y \rightarrow  \bar V  \rightarrow \bar U \rightarrow \bar X~.  
\end{equation} 

In this way  we obtain $\bar X$, i.e., the linear combination 
of the columns of the Jacobian of the function $X \to Y$,
with weights given by the input $ \bar Y$ (e.g., $1,0,\ldots,0$), namely,
\begin{equation}\label{adjoint}
\bar X_i = \sum_{j=1}^m \bar Y_j \frac{\partial Y_j}{\partial X_i} ~,
\end{equation}
with $i=1,\ldots, n$.

In the adjoint mode, the cost does not increase with the number of inputs, but it is linear in the number
of (linear combinations of the) columns of the Jacobian that need to be evaluated independently. In particular, if the full Jacobian is
required, one needs to repeat the adjoint calculation  $m$ times, setting the vector $\bar Y$ equal to each of the
elements of the canonical basis in $\mathbb{R}^m$.
Furthermore, since  the partial derivatives depend on the values of the intermediate
variables, one generally first has to compute the original calculation storing the values of
all of the intermediate variables such as $U$ and $V$, before performing the adjoint mode
sensitivity calculation.  

One particularly important theoretical result is that given a computer code 
performing some high-level function (\ref{function}), the execution time of its adjoint counterpart
\begin{equation}\label{adjointfunction}
\bar X = \texttt{FUNCTION}\_{\texttt B}(X, \bar Y)
\end{equation}
(with suffix $\_{\texttt B}$ for ``backward'') calculating 
the linear combination (\ref{adjoint}) is bounded by approximatively 4 times the cost of
execution of the original one.
Thus, one can obtain the sensitivity of a single output, or of a linear combination of outputs, 
to an unlimited number of inputs for little more work than the original computation.

The propagation of the adjoints, being mechanical in nature can be automated. Indeed,
several AD tools \cite{autodiff} are available that, given a
function of the form (\ref{function}), generate the adjoint function (\ref{adjointfunction}).
While the application of such automatic AD tools to
large inhomogeneous simulation software is challenging, the
principles of AD can be used as a programming paradigm
of any algorithm.
This is especially useful for the most common situations where
simulation codes use a variety of libraries written in different
languages, possibly linked dynamically.
However, automatic tools are of great utility to generate the adjoint of self contained 
functions and subroutines thus effectively reducing the development time of adjoint implementations.

A detailed tutorial on the programming techniques that are useful for adjoint implementations is 
beyond the scope of this paper. However, when hand-coding the adjoint counterpart of a 
set of instructions in a general algorithm it is often enough to keep in mind just a few 
practical recipes, for instance:

\begin{enumerate}

\item[i)] As previously mentioned, each intermediate differentiable variable $U$ can be used not only 
by the subsequent instance but also by several others occurring later in the 
program.  
As a result, the adjoint of $U$ has 
in general several contributions, one for each instruction of the original function in which $U$ was on the 
right hand side of the equal sign (assignment operator). Hence, by exploiting the linearity of differential operators, 
it is in general easier to program according to a syntactic paradigm in which adjoints are always updated 
so that the adjoint of an instruction of the form 
$$
V = V(U)
$$
reads
$$
\bar U_i = \bar U_i + \frac{\partial V_k(U)}{\partial U_i} \bar V_k~. 
$$
Clearly, this implies that the 
adjoints have to be appropriately initialized as discussed in the following
paragraphs.
In particular, to cope with input variables that are
changed by the algorithm  (see next point) it is generally best to initialize to zero the adjoint of a given variable in correspondence of 
the instruction in which it picks its first contribution (i.e., right before the adjoint corresponding to  the last instruction of the original code 
in which the variable was to the right of the assignment operator).
 For instance, the adjoint of the following sequence of instructions
\begin{eqnarray*}
 y &=& F(x)\\
 z &=& H(x,y)\\
 x &=& G(z)
 \end{eqnarray*}
 can be written as:
 \begin{eqnarray*}
 \bar z &=& 0 \\
 \bar z &=& \bar z +\frac{\partial G(z)}{\partial z} \bar x \\
 \bar y &=& 0 \\
 \bar x &=& 0 \\
 \bar x &=& \bar x + \frac{\partial H(x,y)}{\partial x} \bar z \\
 \bar y &=& \bar y +\frac{\partial H(x,y)}{\partial y} \bar z \\
 \bar x &=& \bar x + \frac{\partial F(x)}{\partial x} \bar y~.
 \end{eqnarray*}
 Note in particular that the symbol $x$ represents the input and the output 
of the algorithm. As a result, also  
$\bar x$ represents  both the  input and the output of the adjoint algorithm 
and it is crucial to reinitialize to zero $\bar x$
before it  picks its  first contribution from an adjoint statement, i.e., the one associated with the instruction $z=H(x,y)$.
As explained in detail in the following example, the algorithm could be more
 easily understood by replacing the last statement
by another independent 
output variable $u=G(z)$, and following the straightforward 
derivation of the adjoint algorithm that has for input $\bar u$ and output 
$\bar x$ (see also the example below). One can easily derive that the resulting algorithm 
 coincides with the one above, namely the same input  provides the same output. 

\item[ii)] In some situations the input $U$ of a function $V=V(U)$ is modified by the function itself. This situation is easily 
analyzed by introducing an auxiliary variable $U^\prime$ representing the value of the input after the function evaluation. As a result,
the original function can be thought of the form $(V,U^\prime) = (V(U), U^\prime(U))$, where $V(U)$ and $U^\prime(U)$ do not mutate their inputs, in combination with the assignment $U=U^\prime$, overwriting the
original input $U$. The adjoint of this pair of instructions clearly reads
\begin{eqnarray*}
 \bar U^\prime_i &=& 0\\
 \bar U^\prime_i &=& \bar U^\prime_i + \bar U_i~,
\end{eqnarray*}
where we have used the fact that the auxiliary variable $U^\prime$ is not used elsewhere (so $\bar U^\prime_i$ does not have any previous contribution), and
 \begin{eqnarray*}
 \bar U_i &=& 0\\
 \bar U_i &=&  \bar U_i+ \frac{\partial V_k(U)}{\partial U_i} \bar V_k + \frac{\partial U^\prime_l(U)}{\partial U_i} \bar U^\prime_l~.
 \end{eqnarray*}
 where, again, we have used the fact that also the original input $U$ is not used after the instruction $V=V(U)$, as it gets overwritten.
 One can therefore eliminate altogether the adjoint of the auxiliary variable $\bar U^\prime$ and simply write
 \begin{equation*}
 \bar U_i =  \frac{\partial V_k(U)}{\partial U_i} \bar V_k + \frac{\partial U^\prime_l(U)}{\partial U_i} \bar U_l~.
 \end{equation*}
Very common examples of this situation are given by increments of the form
$$
U_i = a\, U_i + b
$$
with $a$ and $b$ constant with respect to $U$. According to the recipe above, the adjoint counterpart of this instruction simply reads
$$
\bar U_i =  a\, \bar U_i~. 
$$
These situations are common in iterative loops where a number of variables are typically updated at each iteration. 

%
\end{enumerate}

\begin{figure}[!ht]
\vspace{5cm}
\includegraphics[width=12cm,angle=90]{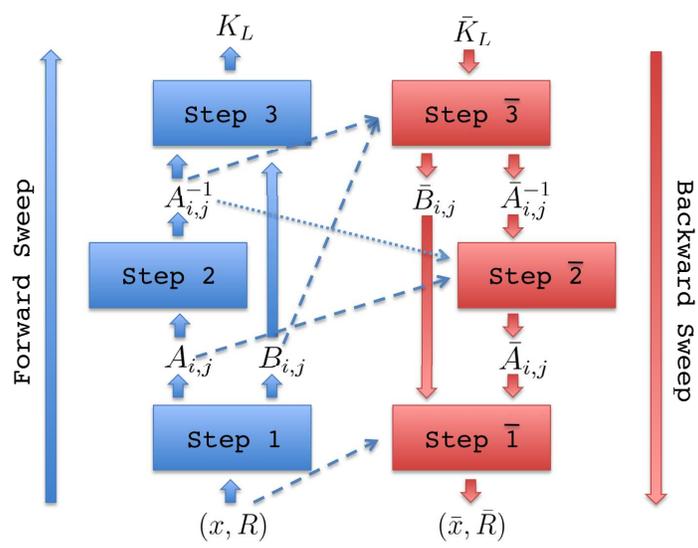}
\vspace{5cm}
\caption{\label{aad_diagram} S. Sorella and L. Capriotti, Journal of Chemical 
Physics}
\end{figure}

In order to better illustrate these ideas, here we consider the calculation of the kinetic energy
and of its adjoint counterpart, where for simplicity we consider only one spin 
component (maximum polarized case), as the calculation of both spin up and 
spin down contributions can be obtained just by summing them. 
Given the position of the electrons $x$ and of the ions $R$, 
the calculation of the kinetic energy can be performed according to the following steps, as derived in App.~\ref{laplacian}:
\begin{enumerate}

\item Calculate $A_{i,j} = \psi_i(\mathbf{r}_j)$ and $B_{i,j} = \Delta_j \psi_i(\mathbf{r}_j)$
according to the definition of the molecular orbitals in Eq.~(\ref{defpsi}). 

\item Calculate $A^{-1}_{i,j}$ by matrix inversion.

\item Calculate the kinetic energy as 
\begin{equation} \label{lapeq}
K = -{1 \over 2} 
 \sum_{i,j} A^{-1}_{i,j} B_{j,i}.
\end{equation}

\end{enumerate} 
The corresponding adjoint algorithm can be constructed by associating to each of the steps
above its adjoint counterpart according to the correspondence given by 
Eqs.~(\ref{function}), (\ref{adjoint}), 
and ~(\ref{adjointfunction}).  As a result, as also illustrated schematically in Fig.~\ref{aad_diagram}, 
the adjoint algorithm for the derivatives of the kinetic energy with respect to  the positions of 
the electrons and ions, consists of steps 1-3 above, and their adjoint counterparts 
executed in reverse order, namely:
\begin{enumerate}

\item[$\bar{3}$.] Set $\bar K = 1$, and evaluate the adjoint of the function $(A^{-1}_{i,j}, B_{j,i}) \to K$ defined in step 3. 
This is a function of the form $(A_{i,j}^{-1}, B_{i,j},\bar K) \to (\bar A_{i,j}^{-1}, \bar B_{i,j})$  with 
$\bar A^{-1}_{i,j} = -{1 \over 2} \bar K B_{j,i} $ and $\bar B_{i,j} =-\frac{1}{2} \bar K A^{-1}_{j,i}$. 

\item[$\bar{2}$.] Evaluate the adjoint of the function $A_{i,j} \to A^{-1}_{i,j}$ (step 2), namely,
$(A_{i,j}, \bar A^{-1}_{i,j}) \to \bar A_{i,j}$ with $\bar A = -(A^{-1})^T \bar A^{-1} (A^{-1})^T$ (see App. \ref{mathinv}).

\item[$\bar{1}$.] Evaluate the adjoint of the function $(x,R) \to (A_{i,j}, B_{i,j})$ (step 1), namely,
$(x, R, \bar A_{i,j}, \bar B_{i,j}) \to (\bar x, \bar R)$ with
\begin{eqnarray*}
\bar x_j &=& {  \partial K  \over \partial {\mathbf{ r}_j} } =\sum_{i} \bar A_{i,j} \partial_{\mathbf{ r}_j}  \psi_i(\mathbf{r}_j) + \bar B_{i,j} \partial_{\mathbf{ r}_j}  \Delta_j \psi_i(\mathbf{r}_j)~, \\
\bar R_a &=&  {  \partial K  \over \partial {\mathbf{ R}_a} } = \sum_{i,j}  \left[ \bar A_{i,j} \partial_{\mathbf{ R}_a}  \psi_i(\mathbf{r}_j) + \bar B_{i,j} \partial_{\mathbf{R}_a}  \Delta_j \psi_i(\mathbf{r}_j) \right]  \\ 
   &=& \sum_{i,j,k} \chi_{i,k}  \delta_{\mathbf{ R}_k,\mathbf{ R}_a} 
 \left[ \bar A_{i,j} \partial_{\mathbf{ R}_k}  \phi_k(\mathbf{r}_j) + \bar B_{i,j} \partial_{\mathbf{R}_k}  \Delta_j \phi_k(\mathbf{r}_j) \right]~,
\end{eqnarray*}
where in the latter equality we have expanded the orbitals 
in terms of atomic orbitals, by means of  Eq.(\ref{defpsi}).
\end{enumerate} 
Notice that in the last expression it is the presence of the Kronecker delta, that allows the computation of all the derivatives with respect to the
atomic positions $ {\mathbf R}_a$ in $ \simeq 2 N^2 L$ operations, namely the 
same amount of operations used in the forward step.
Indeed  by summing only {\rm once} 
over the three indices $i,j,k$ in the above expression all the force components 
acting on all the  atoms are obtained.

In AAD this is not accidental, and the structure of the algorithm is automatically optimized for computing several derivatives at the cheapest computational 
cost.


By applying the chain rule it is immediate to see that $\bar x$ and $\bar R$ computed according to the
steps above are the derivatives of the kinetic energy with respect to the position of the electrons and
the ions, respectively. It is also easy to realize that -- as expected according to general results on the computational
complexity of adjoint algorithms \cite{griewank}
quoted above -- the number of operations involved in each adjoint step is a small constant times the number of operations of the original step, namely (considering only multiplications) $2 N^2 L$ vs $N^2 L $, $2 N^3$ vs $N^3$, and $2N^2$ vs
$N^2$ for steps $\bar{1}$ vs 1, $\bar{2}$ vs 2, and $\bar{3}$ vs 3, respectively.  

As also anticipated, the propagations of the adjoints (steps $\bar{3}-\bar{1}$) can be performed only after the calculation of 
the kinetic energy has been completed (steps $1-3$) and some of the intermediate results (e.g., the matrices $A$, 
$B$, and $A^{-1}$) have been computed and stored. This is the reason why, in general, the adjoint of a given function 
generally contains a {\em forward sweep}, reproducing the steps of the original function, plus a {\em backward sweep}, 
propagating the adjoints.  This construction can be clearly applied recursively for each of the steps involved in the 
calculation. 

It is worth noting that each adjoint step, taken in isolation, contains in turn a forward sweep, recovering the 
information computed in the original step that is necessary for the propagation of the adjoints.  
However, this can be clearly avoided by storing such information at the time it is first computed in the original step.
Strictly speaking, this is necessary to ensure that the computational cost of the overall algorithm remains 
within the expected bounds. However, there is clearly a tradeoff between the time necessary to store and retrieve 
this information and 
the time to recalculate it from scratch, so that in practice it is often enough to store in the main forward sweep only the 
results of relatively expensive computations. In the example above for instance, significant savings can be obtained by storing
the inverse of the matrix $A$ at the output of step 2 and passing it as an input of Step $\bar 2$ (see Fig.~\ref{aad_diagram}).

The main complication in the algorithm above is the implementation of the adjoint of the Laplacian of the WF 
in step $\bar{1}$. However, the calculation of the Laplacian is a good example of an instance that can be represented by 
a self contained, albeit complex, computer function, for which several automatic differentiation tools are available. 
In particular, in order to complete step $\bar{1}$, it is enough to define the 
adjoint functions  of the calculation of the Laplacian 
$ \vec r \to \Delta  \phi_j (\vec r) $, 
for a set of explicit functions $\{\phi_j \} $ 
(e.g. gaussians). 
The adjoints are then computed by means of  the corresponding 
gradient of the Laplacian, namely $ \bar \psi \to \vec{ \bar  r}  $ 
where $  \vec{ \bar r} = \vec {\bar r} + \bar\psi \nabla_{\vec r } \Delta \phi_j ( \vec r)$. 
For this application  we have used TAPENADE, 
developed at INRIA by Hasco\"et and collaborators \cite{tapenade}.

\section{Results}
\begin{figure}[!ht]
\vspace{5cm}
\includegraphics[width=10cm]{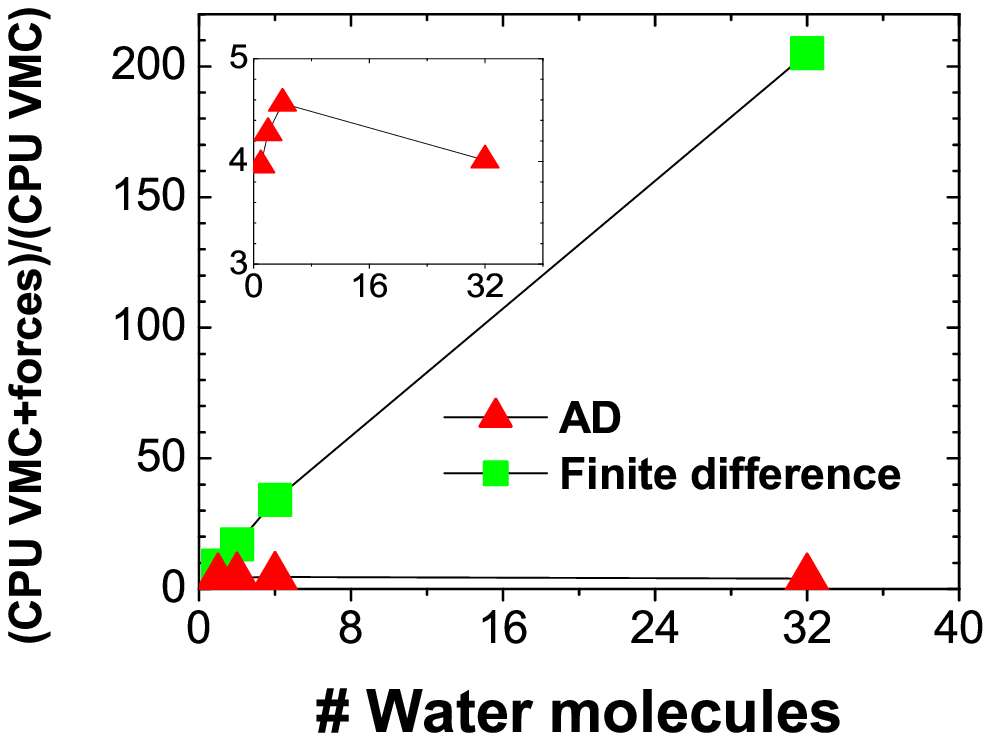}
\vspace{5cm}
\caption{\label{adcpu} S. Sorella and L. Capriotti, Journal of Chemical 
Physics}
\end{figure}

After implementing the adjoint counterpart of the two main instances 
corresponding to the evaluation of the log WF and the local energy, 
we have computed 
the exact energy derivatives and compared with the straightforward finite-difference evaluation, finding perfect agreement within numerical accuracy.
However, the finite-difference method presents a well known bottleneck:
in order to evaluate the $3M$ energy derivatives, one has to evaluate the local 
energy and the log WF at least $3 M $ times more.
Since the computation of such quantities is the most relevant part in 
QMC, with a computational effort 
scaling as $N^3$, we end up with a very inefficient  algorithm  for large 
number of atoms. 
As shown in Fig.~\ref{adcpu}, this slowing down can be completely removed 
by using AAD, as the cost to compute 
all the force components in a system containing several water molecules, 
remains approximately $4$ times larger than the cost to compute only the 
total energy. 
This factor $4$ is a  very small cost, if we consider that the 
main adjoint instance has to be evaluated twice, one for the local energy 
and the other for the WF logarithm, and that, on the other hand,  
VMC is the 
fastest method in QMC. For instance, we can evaluate forces within LRDMC 
with only a small overhead, 
as the cost to generate a new independent configuration 
within LRDMC is about 10 times larger than VMC, and therefore, for this 
more accurate method, the 
cost to compute all force components will be essentially negligible.
Analogous consideration holds during an energy optimization. 
We have to consider that in this case AAD can be used to compute not only 
the force components, but also all the energy derivatives with respect to 
all variational parameters $ \{ c_i \} $ of the WF, 
essentially at the same computational cost, even when the number 
$p$ of variational parameters is extremely large.

Though we have not implemented AAD for this general task, we expect a further 
speed up (and simplification) of the code, once AAD  will be fully implemented 
for all possible energy derivatives. We believe this will become common 
practice for future quantum Monte Carlo packages.  
At present, in order  to have consistent forces within VMC,  
all variational parameters have to be 
optimized \cite{rappe}, 
and to this purpose we have used the standard way to compute 
energy derivatives.  

We have applied the efficient evaluation of the forces for the structural 
optimization of the water monomer.
We have used energy-consistent pseudopotentials \cite{filippipseudo}  only for the oxygen atom. 
In the calculation we have adopted a huge basis set to avoid basis 
superposition errors.
The molecular orbitals are expanded in a primitive basis containing 
24s22p10d6f1g on the oxygen and 6s5p1d on the hydrogen atom. 
The exponents of the Gaussians are optimized by minimizing the 
energy of a self-consistent DFT calculation within the LDA approximation \cite{azadi}.
The accuracy in the total DFT energy is well below $1mHa$ for the water dimer, 
implying  that we are essentially working with an almost complete basis set.  
 For the Jastrow factor  we have also used a quite large basis, to achieve 
similar accuracy in the total energy, within a VMC calculation on a WF 
 obtained by  
optimizing the Jastrow over the LDA Slater determinant. 
The final optimized basis for the Jastrow 
contains  a contracted basis 6s5p2d/3s3p1d on 
the oxygen and an uncontracted 1s1p basis on the hydrogen atom.

\begin{figure}[!ht] 
\vspace{5cm}
\includegraphics[width=10cm]{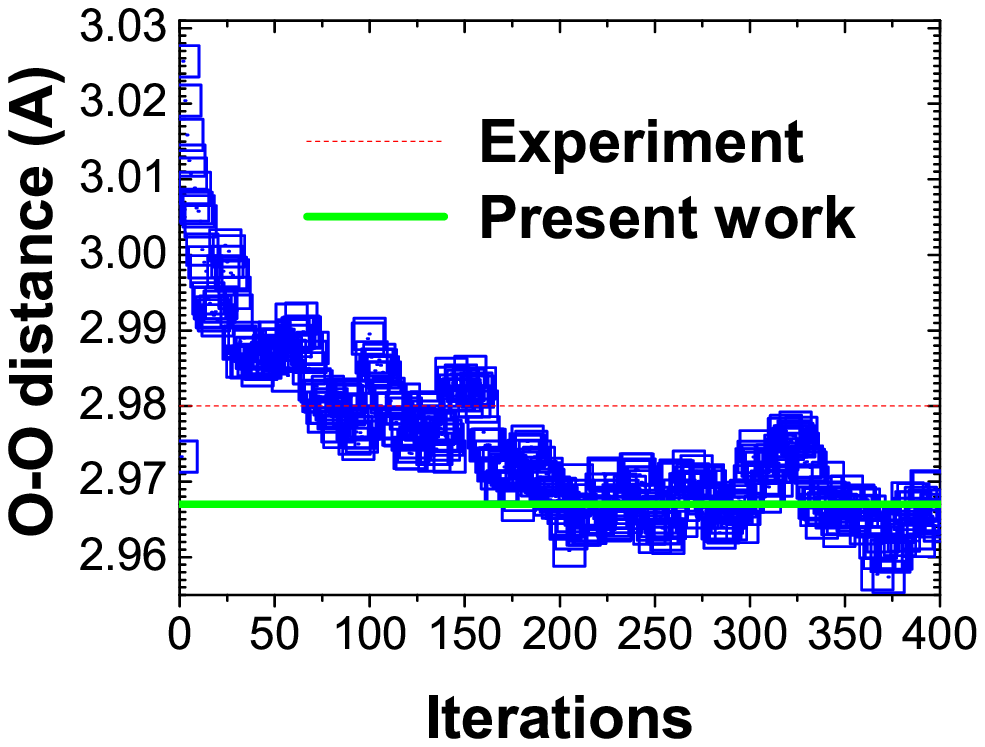}
\vspace{5cm}
\caption{\label{dimer} S. Sorella and L. Capriotti, Journal of Chemical 
Physics}
\end{figure}

In the following we describe the first application of this method 
for optimizing the structure of simple water compounds.
 The variational parameters of the 
WF -- molecular orbitals and Jastrow factor -- are 
optimized, by energy minimization, with the method 
described in Ref.~\onlinecite{marchi}.
At each step of optimization, we compute the ionic forces by AAD,
 and we employ a standard steepest descent 
move of the ions  $ \mathbf{ R}_a \to \mathbf{ R}^\prime_a$: 
\begin{equation}
\mathbf{ R}^\prime_a= \mathbf{ R}_a + \Delta \tau \mathbf{ F}_a 
\end{equation} 
where $\Delta \tau = 1/2 a.u.$.
After several hundred iterations both the variational parameters and 
the atomic positions fluctuate around average values, and we use the last 
few hundred iterations to evaluate the error bars and the mean value 
of the atomic positions, as illustrated in Fig.~\ref{dimer}.

In Tab.~\ref{watermon} we show the optimized structure of the water monomer. 
As it is clearly evident our final atomic positions are almost indistinguishable from the experimental ones. Generally speaking our calculation appears more 
accurate than simple mean field 
 DFT methods, and comparable with state of the art 
quantum chemistry techniques, such as CCSD(T). 
The accuracy of the VMC method has been also confirmed recently in another context.\cite{valsson}

In the dimer structure the situation is slightly different.
As shown in Tab.~\ref{waterdim}, the oxygen-oxygen distance is in quite 
good agreement with experiments, whereas the $OHO$ angle is overestimated by 
few degrees. Probably in this case the quantum corrections
should affect  the 
hydrogen position between the two oxygens, because the 
dimer bond is very weak.
Indeed we have also checked that,  with the more 
accurate LRDMC calculation,  the equilibrium structure obtained by the 
VMC method remains stable as all the force components are well below 
$10^{-3} a.u.$. On the other hand LRDMC increases
 the binding of the dimer by about $  1 Kcal/mol$, showing that, from the energetic point of view, the LRDMC calculation may be 
important, as also confirmed in previous studies \cite{marchi,needswat}.  
All the above calculations can be done with a relatively small computational 
effort (few hours in a 32 processor parallel computer), and therefore the 
same type of calculation, with the same level of accuracy, can be extended 
to much larger systems containing several atoms 
with modern supercomputers.

\begin{figure}[!ht]
\vspace{5cm}
\includegraphics[width=10cm]{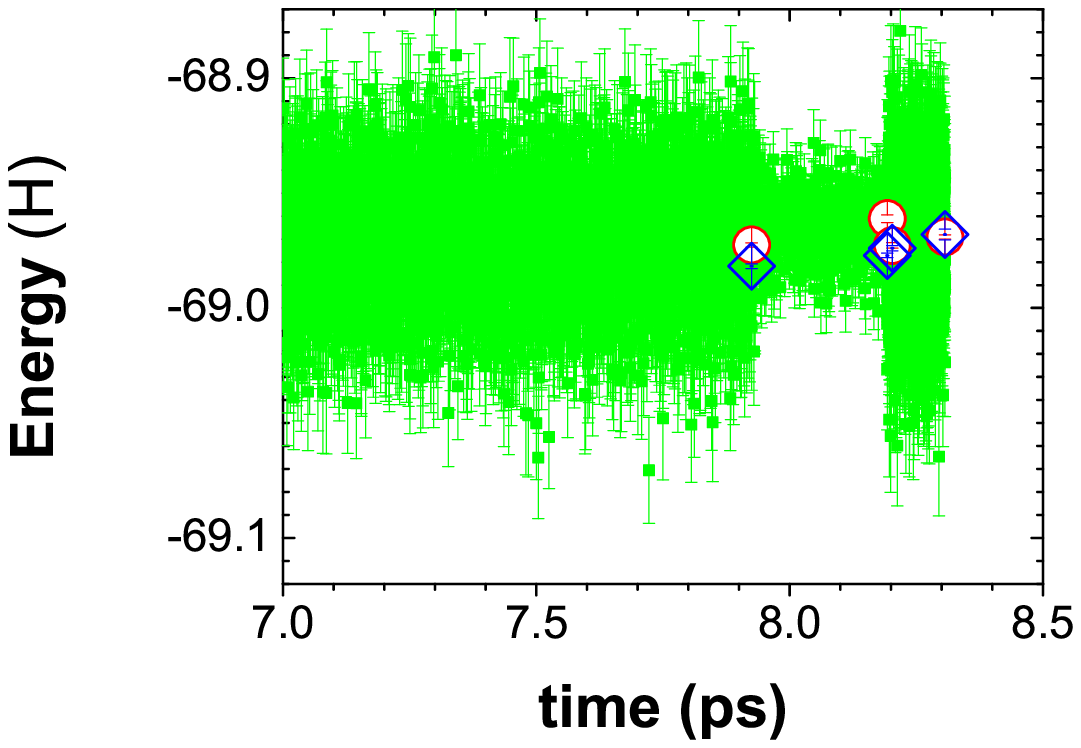}
\vspace{5cm}
\caption{\label{energy} S. Sorella and L. Capriotti, Journal of Chemical 
Physics}
\end{figure}

Stimulated by the above success we have tested the finite-temperature 
molecular dynamics simulation introduced some time ago \cite{attacc}, using 
4 water molecules in a cubic box with $4.93$\AA ~side length,  mimicking
 the density
of liquid water at ambient conditions. Since we are interested in 
static equilibrium properties we have used for the oxygen the same mass of hydrogen.
Though the system is very small we have been able to perform several thousands 
steps.  For  each step all variational parameters are optimized using 
a given number $n$ of stochastic reconfiguration (SR) optimizations \cite{rocca}.
For the  first $18000$ 
steps we used $n=1$ and a  time integration step $\Delta t$ 
 for the MD ranging from  $20 a.u.$
or $40 a.u.$. 
Several iterations were possible because in QMC we can decide to work 
with a relatively small number of samples to accumulate statistics for the 
energy derivatives and the forces. In these conditions the forces are rather 
noisy but the molecular dynamics with noise correction \cite{attacc}
 allows us to have sensible 
results, at the price to have  an  overdamped dynamics.
However, as it is shown in Fig.~\ref{energy}, it is difficult to remain 
within the Born-Oppheneimer energy surface, because at selected times, we 
have fully optimized the wave-function using further 200 SR iterations, and 
found 6$mHa$ difference between the energy on fly and the optimized energy.
In order to overcome this bias in the dynamics, in the final part of the 
MD simulation, we have used $n=10$ (and $\Delta t=40 a.u.$) 
and found that we remain sufficiently 
close to the Born-Oppheneimer energy surface. 
This very preliminary application is clearly limited by 
the too small number of water molecules considered in the simulation, 
and therefore does not allow us to 
determine the equilibrium 
properties of liquid water. Nevertheless, we believe that this result is  
rather encouraging because it shows that all 
the possible sources of errors in the MD driven by QMC forces, can be 
controlled in a rather straightforward way.

\section{Conclusions}
In this work we have shown that the calculation of all the force components 
in an electronic system containing several atoms, can be done very 
efficiently using adjoint algorithmic differentiation (AAD).
In particular it is possible to employ the very efficient space warp coordinate transformation (SWCT) in differential form in 
a straightforward and simple way, even when pseudopotentials and/or complicated 
many-body wave functions are used.
More importantly, we have shown that, using AAD, one can compute all these force components, and in principle 
all the energy derivatives with respect to any variational parameter contained 
in the many-body wave function, in about four times the cost 
to compute the expectation value of the energy. 

So far, for large number of atoms, 
the use of quantum Monte Carlo methods have been generally limited to total energy calculations. We believe that our work opens the way for new and more 
accurate tools for ab-initio electronic simulation based on quantum Monte 
Carlo.
In particular we have shown that it is possible to perform an ab-initio 
molecular dynamics simulation for several picoseconds, in a system containing
four water molecules. Since the cost of a variational Monte Carlo 
calculation with fixed statistical accuracy in the energy per atom (total  
energy) increases with the number of atoms as $M^2$ ($M^4$) the simulation 
of  about $32$ water molecules should be possible with less than $10^5$ 
($10^7$) CPU hours, a figure that is nowadays possible (at the limits of 
present possibilities)  with modern massively parallel supercomputers.
It is not known at present if
it is sufficient to target a fixed statistical error in the energy per atom 
in order to obtain well-converged thermodynamic extensive 
quantities. Otherwise a computationally more expensive calculation 
with a statistical error on the total energy of the order of $kT$ is necessary, as in the penalty method\cite{ceperleypenalty}.
  
In the example we have presented, we have also seen that the 
accuracy of variational Monte Carlo in determining the equilibrium structure of the water 
monomer and the water dimer is rather remarkable and comparable to post 
Hartree-Fock methods,  requiring much more  computer resources for 
large number of atoms.
Therefore we believe that, in view of the efficiency in the evaluation 
of forces  obtained by AAD, realistic and very accurate ab-initio simulation
based on quantum Monte Carlo will be within reach in the near future. 

\acknowledgements

This work was partially supported by COFIN2007, and CNR. 

\appendix
\section{ Calculation of the Laplacian }
\label{laplacian}
In order to compute the Laplacian $K$ 
of a Slater determinant WF:
\begin{equation}
\langle x | S \rangle = \det A
\end{equation} 
where $ A $ is the $N \times N$ matrix defined by:
\begin{equation}
 A_{i,j} = \psi_i (\mathbf{ r_j } )
\end{equation}
and the molecular orbitals $\psi_i$ are defined in Eq.~(\ref{defpsi}) and 
the spin index is omitted for simplicity.

We notice that if we change the electron position of the $k^{th}$ electron
the matrix $A$  change only by a single column:
\begin{equation}
 A^\prime_{i,j} = A_{i,j} + \delta_{j,k} \left[ \psi_i (\mathbf{ r}^\prime_k) - 
A_{i,k} \right]
\end{equation}
In this way we can single out the dependence on $ \mathbf{ r}^\prime_k $ 
by evaluating explicitly the ratio of the two WF's:

\begin{equation}
{ \det A^\prime \over \det A } =\det A^{-1} A^{\prime} = \sum_j A^{-1}_{k,j} 
\psi_j(  \mathbf{ r}^\prime_k ) 
\end{equation}
The evaluation of the Laplacian with respect to the $k^{th}$ electrons 
easily follows by linearity of differential operations 
such as the Laplacian, so that we finally arrive to  
Eq.~(\ref{lapeq}), by summing over all the electrons.

\section{ Calculation of the adjoint of an inverse matrix operation 
}
\label{mathinv}
In this appendix we derive the adjoint instance of the 
calculation of the inverse $A^{-1}_{ij}$ of an input  $N \times N$ 
square matrix $A_{i,j}$.
To this purpose we notice that  
an arbitrary linear combination of the output (the inverse matrix) 
can be written as:
\begin{equation} \label{eqinv}
\sum\limits_{i,j} \bar A^{-1}_{i,j} A^{-1}_{i,j} =
 {\rm Tr} \left[ ( \bar A^{-1} )^T A^{-1} \right]
\end{equation}
where the subscript  $T$ indicate the transpose of the corresponding 
matrix, and ${\rm Tr}$ the conventional matrix trace (sum of the diagonal elements).
As explained in Sec.~\ref{adsec} by differentiating the above equation with respect to an arbitrary 
variation of the input we obtain the adjoint instance.
To this purpose we denote  
the  differential change of the 
input $A$ with the matrix $DA$, 
so that the corresponding variation can be conveniently written as: 
$$ (A + D A)= A ( I + A^{-1} DA) $$
and therefore :
$$ (A+ DA)^{-1}= ( I - A^{-1} DA) A^{-1} + O (|DA|^2). $$
Thus, by simply substituting the above relation  in Eq.~(\ref{eqinv}), 
and by using the cyclic invariance of the trace,  we 
obtain:
\begin{equation}
d {\rm Tr} \left[ ( \bar A^{-1} )^T A^{-1} \right]
= -{\rm Tr} \left[ A^{-1} (\bar A^{-1})^T) A^{-1} DA \right]
\end{equation}
which implies that the adjoint matrix $\bar A$, after this instance, is 
updated as follows:
\begin{equation}
\bar A = \bar A+{  d {\rm Tr} \left[ ( \bar A^{-1} )^T A^{-1} \right] \over d A }
= \bar A - \left[ A^{-1} (\bar A^{-1})^T) A^{-1} \right]^T = \bar A
-(A^{-1})^T \bar A^{-1} (A^{-1})^T 
\end{equation} 
which concludes this appendix.

%
%
\newpage
\begin{table}[!hbp]
\begin{center}
\caption{ Efficiency of the various types of evaluation of forces in VMC for 
a water dimer molecule at experimental equilibrium atomic positions.
The computational time to calculate the force components on the Oxygen atoms 
within a given statistical 
error is inversely proportional to the efficiency. The HFM efficiency is 
taken for reference equal to one, for a statistical accuracy of 0.001a.u.. 
Pseudopotential is used only for the Oxygen.
}
\label{tabeff}
\begin{tabular}{ |c | c| c| c|  }
 \hline
 Method  &  HFM  & No-SWCT  &  SWCT  \\
\hline
 Efficiency & 1 & 165 & 1360\\
\hline 
\end{tabular}
\end{center}
\end{table}
\newpage
\begin{table}[!hbp]
\begin{center}
\vspace{5cm}
\begin{threeparttable}
\caption{ VMC optimized structure of the water monomer. }
\label{watermon}
\begin{tabular}{ |c | c| c| c| c| c| c| }
\hline\hline
  & Exp  & VMC  & LDA\tnote{a}~~   & BLYP\tnote{a}~~  & BP\tnote{a}~~ 
 & CCSD(T) \tnote{b}~~  \\
\hline
$d_{OH}(A) $  &0.957\tnote{c}~~   & 0.954(1) &  0.973  &  0.973   &   0.974   & 0.95829\ \\
\hline
$\angle HOH$ (deg)  & 104.5\tnote{d}~~ & 104.61(10)  &  104.4 & 104.6 & 104.1  & 104.454 \\
\hline\hline
\end{tabular}

\begin{tablenotes}
{\footnotesize
\item[a]{From Ref.~\cite{sprik}}
\item[b]{From Ref.~\cite{ccsdt}}
\item[c]{From Ref.~\cite{expmon1}}
\item[d]{From Ref.~\cite{expmon2}}
}
\end{tablenotes}
\end{threeparttable}
\vspace{5cm}
\end{center}
\end{table}

\newpage
\begin{table}[!hbp]
\begin{center}
\vspace{5cm}
\begin{threeparttable}
\caption{ VMC optimized structure of the water dimer. The LRDMC calculation 
was done only for the binding energy, by projecting the VMC WF.}
\label{waterdim}
\begin{tabular}{ |c | c| c| c| c| c| c| c|  }
\hline\hline
  & Exp  & VMC  & LRDMC & LDA\tnote{a}~~   & BLYP\tnote{a}~~  & PBE\tnote{b}~~ 
  & CCSD(T) \tnote{c}~~  \\
\hline
$d_{OO}(A) $  &2.98\tnote{d}~~   & 2.969(2) & =  &  2.70  & 2.95  &   2.89   & 2.9089  \\
\hline
$\angle OHO$ (deg)  & 174\tnote{e}~~ & 177.6(3)  & = & 169  & 173 & 172  & - \\
\hline\hline
binding (kCal/mol)  & 5.0(7) \tnote{f}~~ & 3.84(14)  & 4.76(6) &  8.8 & 4.3 & 5.55  & = \\
\hline\hline
\end{tabular}

\begin{tablenotes}
{\footnotesize
\item[a]{From Ref.~\cite{sprik}}
\item[b]{From Ref.~\cite{pbe}}
\item[c]{From Ref.~\cite{ccsdtdim}}
\item[d]{From Ref.~\cite{expdim1}}
\item[e]{From Ref.~\cite{expdim2}}
\item[f]{From Ref.~\cite{expdim3}}
}
\end{tablenotes}
\end{threeparttable}
\vspace{5cm}
\end{center}
\end{table}

\newpage

List of figures

\begin{itemize}
\item
Figure 1:
Shematic representation of the adjoint algorithm for the calculation of the local kinetic energy. Note that the inverse of the matrix $A$
can be passed directly as an input (dotted arrow) of step $\bar 2$ thus avoiding to repeat the matrix inversion.

\item
Figure 2:
Ratio of CPU time required to compute energies and all force 
components referenced to the one required for the 
simple energy calculation within VMC.
The calculations refer to $1,2,4$, and 32 water molecules.
The inset is an expansion of the lower part of the plot.

\item
Figure 3:
oxygen-oxygen distance as a function of the number of iterations for determining the equilibrium zero-temperature structure of the water dimer. 
All the 18 atomic coordinates, as well as about 1000 variational 
parameters of the electronic many-body WF are  fully optimized with an iterative scheme \cite{rocca,marchi}. 


\item
Figure 4:
Average internal energy  as a function of the simulation time, for the 
molecular 
dynamics with QMC forces\cite{attacc} and  four water 
molecules in a cubic box.
Empty dots indicate VMC energy values corresponding to the 
time evolved WF with much smaller 
error bars ($<0.4mHa$). 
Empty squares are obtained after optimizing the WF at 
fixed atomic positions, starting from the previous initial state  and 
with quite accurate statistical accuracy ($<0.4mHa$).

\end{itemize}

\newpage

\newpage

\newpage

\end{document}